\documentclass[%
aip,
sd,%
amsmath,amssymb,
author-numerical,%
]{revtex4-1}

\def \bF {\pmb{F}}
\def \bH {\pmb{H}}

\def \bx {\pmb{x}}

\def \bs {\pmb{s}}

\usepackage{graphicx}
\usepackage{dcolumn}
\usepackage{bm}
\usepackage[dvipsnames]{xcolor}
\usepackage[normalem]{ulem}
\usepackage[title]{appendix}
\usepackage{comment}
\usepackage{float}
\usepackage{mathtools}
\usepackage{amsthm}
\usepackage{hyperref}
\usepackage{mathtools}
\usepackage{url}
\usepackage{natbib}

\graphicspath{{./figures/}}
\newtheorem{lemma}{Lemma}
\newtheorem{definition}{Definition}
\newtheorem{remark}{Remark}
\begin{document}
		
			\title{\bf {Cluster Synchronization of Networks via a Canonical Transformation for  Simultaneous Block Diagonalization of Matrices}}
	
\author{Shirin Panahi}
\affiliation{ 
	Mechanical Engineering Department, University of New Mexico, Albuquerque, NM 87131}

\author{Isaac Klickstein}
	\affiliation{ 
		Mechanical Engineering Department, University of New Mexico, Albuquerque, NM 87131}

\author{Francesco Sorrentino}
	\email{fsorrent@unm.edu}
	\affiliation{
		Mechanical Engineering Department, University of New Mexico, Albuquerque, NM 87131}

	\begin{abstract}
We study cluster synchronization of networks and propose a canonical transformation for  simultaneous block diagonalization of matrices
that we use to analyze stability of the cluster synchronous  solution. Our approach has several advantages as it allows us to: (1) decouple the stability problem into subproblems of minimal dimensionality while preserving physically meaningful information; (2) study stability of both orbital and equitable partitions of the network nodes and (3) obtain a parametrization of the problem in a small number of parameters. For the last point, we show how the canonical transformation decouples the problem into blocks that preserve key physical properties of the original system.  We also apply our proposed algorithm to analyze several real networks of interest, and we find that it runs faster than alternative algorithms from the literature.\\
{\it Keywords:} Dynamical Network; Simultaneous Block Diagonalization ; Cluster Synchronization.
	\end{abstract}

	\maketitle
	
	\textbf{The problem of cluster synchronization of networks has been studied in a number of papers in the literature, see e.g., \cite{Pecora2014,SA,siddique2018symmetry,Zhang2020}, among others. This paper follows up on these previous studies and proposes a canonical simultaneous block diagonalization routine to decouple the stability problem into subproblems (`blocks') of minimal dimensionality. Our approach has mainly two advantages: (i) each block in which the original problem is decoupled has a clear physical interpretation and (ii) it is faster than other algorithms proposed in the literature when applied to the analysis of real network topologies. It also nicely reconnects with previous work \cite{Pecora2014,SA,siddique2018symmetry} as the results in our paper are found to coincide with those in these other papers, though the techniques used to decouple the stability problem are different.}
	
\section{Introduction} \label{s:intro}

Cluster synchronization (CS) in networks of coupled oscillators
has been the subject of vast research efforts, see e.g., \cite{belykh2008cluster,NSG,dahms2012cluster,fu2014synchronization,kanter2011nonlocal,rosin2013control,williams2013experimental,nicosia2013remote,schaub2016}.
This occurs when the network nodes are divided {into} clusters such that the nodes in each cluster synchronize on the same time evolution but these time evolutions are different for nodes in different clusters.
 Recent work \cite{nicosia2013remote,Pecora2014} has elucidated the relation between
the symmetries of the network topology and the formation of
clusters of synchronized dynamical units in the network. Reference
\cite{Pecora2014} analyzed the formation and stability of synchronized
clusters that correspond to the orbits of the network symmetry
group. References \cite{SA,siddique2018symmetry,Zhang2020} extended this study to the more general case of equitable clusters, where the nodes in each cluster are not necessarily symmetric, but receive the same total input from the nodes in each one of the clusters. Cluster {synchronization} in directed networks was recently addressed in \cite{lodi2021one}.

The master stability function (MSF) approach \cite{Pe:Ca} has been successfully applied to characterize stability of the complete synchronous solution for networks of coupled systems with Laplacian connectivity. The approach is based on decoupling the stability problem into a number of lower-dimensional problems, where each of the lower-dimensional problems depends on an eigenvalue $p_i$ of the Laplacian matrix. The main advantage of this approach is that stability of the {lower-dimensional} system can be parametrized in a generic parameter $p$ and 
one can  determine the range $R$ of the parameter $p$ over which the master stability function $\mathcal{M}(p)<0$. Then stability can be directly assessed, \emph{for any network} of interest, by verifying that the relevant eigenvalues $p_i$ belong to the range $R$, i.e., $p_i \in R$. 
There are, therefore, two main components of this approach: the first one is the dimensionality reduction and the second one is the parametrization. 

The case of cluster synchronization (CS) with non-Laplacian connectivity is more complex. In what follows, we will: (i) reduce the dimensionality of the problem and (ii) obtain a parametrization of the lower dimensional problem in a minimal number of parameters. The novelty of our work lies especially in point (ii). We will see that for the case of CS we can generate a `canonical transformation' of the stability problem that corresponds to a minimal number of parameters.


\section{PROBLEM DEFINITION: Cluster synchronization} \label{Ap2}
A network of coupled dynamical systems can be described by the following set of equations,
\begin{equation}
\dot{\bx}_{i}(t) = \bF(\bx_{i}(t))+\sum_{j=1}^{N}A_{ij}\bH(\bx_{j}(t)) \quad  i=1, \cdots, N
\label{Eq9}
\end{equation}
where $\bx_i(t)$ represents the $m$-dimensional state vector of node $i$ and $\bF : R^{m} \rightarrow R^{m}$ describes the time evolution of each individual system located at node $i$. The adjacency matrix $A$ describes the network connectivity, i.e., $A_{ij}=A_{ji}=1$ if there is a connection between nodes $i$ and $j$ and $A_{ij}=A_{ji}=0$ otherwise. The function $\bH : R^{m} \rightarrow R^{m}$ is the node-to-node coupling function. We call $\mathcal{V}=\{1,...,N\}$ the set of the network nodes.

\begin{definition}\textbf{Equitable cluster partition.}
Given the adjacency matrix $A$, representing the network topology, one can partition the set of the network nodes $\mathcal V$ into subsets that we call equitable clusters, $\mathcal{C}_1,\mathcal{C}_2,..,\mathcal{C}_C$, $\cup_{k=1}^C \mathcal{C}_k=\mathcal{V}$, $\mathcal{C}_k \cap \mathcal{C}_\ell=\emptyset$ for $k \neq \ell$, where 
\begin{equation} \label{ecp}
  \sum_{h \in \mathcal{C}_{\ell}} A_{ih} = \sum_{h \in \mathcal{C}_{\ell}} A_{jh}, \quad \begin{aligned} \forall i,j \in \mathcal{C}_k\\ \forall \mathcal{C}_k,\mathcal{C}_{\ell} \subset \mathcal{V}. \end{aligned}
\end{equation}
\noindent We call $|\mathcal{C}_k|=n_k$ the number of nodes in cluster $k=1,...,C$, $\sum_{k=1}^C n_k=N$.
\end{definition}

In order to find the equitable clusters, we apply the algorithm developed by Belykh and Hasler \cite{belykh2011} to the network with adjacency matrix $A$. The algorithm returns a set of $C$ equitable clusters $\mathcal{C}_1,\mathcal{C}_2,\ldots,\mathcal{C}_C$. Information about each equitable cluster is contained in the $N \times N$ diagonal indicator matrix $E_k=\{E_{k_{ij}}\}$, $k=1,..,C$, where the entry $(i,i)$ of the matrix $E_{k}$ is equal to $1$ if node $i$ is in cluster $\mathcal{C}_k$ and is equal to $0$ otherwise.\\

\begin{definition} \textbf{Colored Network.}
The previously defined equitable clusters induce a colored network, where each node $i$ is assigned a color $k$ if node $i$ is in cluster $\mathcal{C}_k$.
\end{definition}

       Given an equitable partition of the network nodes, we can define an invariant subspace for the set of Eqs.\ \eqref{Eq9}, which we call the cluster synchronization manifold. The dynamics on this manifold 
is the flow-invariant cluster synchronous time evolution \cite{golubitsky2005}  $\lbrace \bs_{1}(t), \bs_{2}(t), \cdots, \bs_{C}(t)\rbrace$, where $\bs_{1}(t)$ is the synchronous solution for all nodes in cluster $\mathcal{C}_1$, $\bs_{2}(t)$ is the synchronous solution for nodes in cluster $\mathcal{C}_2$, and so on.
       
       The $C \times C$ quotient matrix $Q$ is defined such that for each pair of equitable clusters
$\mathcal{C}_k$ and $\mathcal{C}_l$ we have,
\begin{equation}\label{quo1}
    Q_{kl}=\sum_{j \in \mathcal{C}_{l}}A_{ij}  \quad i \in \mathcal{C}_{k}.
    \end{equation}
       The quotient matrix describes a network (the `quotient network'), for which all the nodes in each equitable cluster collapse to a single quotient node. By assuming the system of equations (1) evolves on the cluster synchronization manifold, and averaging over all the nodes in each cluster, we can derive the equations for the time evolution of the quotient network,
     \begin{equation}\label{quo2}
    \dot{\bs}_{k}(t) = \bF(\bs_{k}(t)) + \sum_{l=1}^{C}Q_{kl}\bH(\bs_{v}(t)),
\quad
    k,l = 1, 2, \cdots, C,
\end{equation}
where the $m$-dimensional vector $\bs_{k}(t)$ represents the state of the quotient network node $k=1,...,C$. 

\begin{definition}\textbf{Equitable clusters encoding matrix.} \label{O}
With knowledge of the clusters, we can construct the $N\times C$ equitable clusters encoding 
matrix $O$, 
such that $O_{ij}=1$ if node $i$ is in cluster $\mathcal{C}_j$ and $0$ otherwise for $i=1,\ldots,N$ and $j=1,\ldots,C$.
\end{definition}

\begin{remark}
Given a network described by the adjacency matrix $A$ and the equitable clusters encoding matrix $O$ (see definition \ref{O}), the $C \times C$-dimensional quotient matrix 
$Q$ can be computed as follows \cite{schaub2016},
\begin{equation} 
Q= (O^T O)^{-1} O^T A O.
\end{equation}
\end{remark}


To investigate the stability of the cluster synchronous solution, we consider a small perturbation $\delta\bx_i=(\bx_i - \bs_k)$, $i \in \mathcal{C}_k$. By {linearizing} Eq.\ \eqref{Eq9} about Eq. \eqref{quo2} we obtain the  vectorial equation,
\begin{equation}\label{z}
\delta \dot{\bx}(t) = \left[ \sum_{c=1}^{C}E_{c}\otimes D\bF(\bs_{c}(t))+\sum_{c=1}^{C}A E_{c}\otimes D\bH(\bs_{c}(t))\right] \delta\bx(t),
\end{equation}
in the $mN$- dimensional vector $\delta {\bx}(t)=[\delta {\bx}_1^T(t),\delta {\bx}_2^T(t),...,\delta {\bx}_N^T(t)]^T$. 
We are interested in the possibility that the stability problem for the $mN$-dimensional system Eq.\ \eqref{z} can be decoupled into a set of lower-dimensional equations.
 To this end, we will show how to obtain a transformation 
 matrix $T$ that block-diagonalizes the matrices $A E_c$ while leaving the matrices $E_c$ unchanged. 

\section{Dimensionality Reduction of the Cluster Synchronization Stability Problem} \label{Ap2}

As mentioned in the previous section, our goal is to reduce the {stability} problem into a set of independent lower-dimensional equations instead of dealing with the high-dimensional problem, Eq.\ \eqref{z}. 
Ref.\ \cite{Pecora2014} has used a transformation based on the irreducible representations of the symmetry group to block-diagonalize the set of Eqs.\ \eqref{z} for the case of a network with symmetries. A limitation of this approach is that it is only applicable to the case of `orbital clusters' (corresponding to the orbits of the symmetry group, see e.g. \cite{tinkham2003}) and not to the more general case of `equitable clusters' \cite{kudose2009}. Stability of the cluster synchronous solution in the case of equitable clusters was first discussed in Refs.\ \cite{SA,siddique2018symmetry}. An alternative method based on the approach of simultaneous block diagonalization of matrices (SBD) \cite{Maehara2010-1, Maehara2010-2, Maehara2011}, firstly applied to the problem of complete synchronization of networks in \cite{irving2012synchronization},  was used for cluster synchronization in \cite{Zhang2020}. 
Different from \cite{Zhang2020}, in this paper we propose an SBD transformation that will lead to a `canonical SBD transformation' of the stability problem. Our method applies to both the cases of equitable and orbital clusters. We will discuss the benefits of this canonical transformation in the rest of this paper.


The problem of simultaneous block diagonalization can be formalized as follows: given a set of $N \times N$ symmetric matrices $A^{(1)}, ... , A^{(M)}$, find an $N \times N$ orthogonal matrix $T$ such that the matrices $T^{-1}A^{(k)}T$ have a common block-diagonal structure for $k = 1, ... , M$. It should be noted that such a block-diagonal structure is not unique in at least two senses:  first, the blocks may be permuted, resulting in block diagonal decompositions that are isomorphic; second, the matrices corresponding to certain blocks may be further refined into smaller blocks, resulting in a structures that are fundamentally different. A block diagonal structure with smaller blocks is considered to be finer; we are interested in the finest SBD  as it provides the simplest elements in the decoupling of systems as described above. 

In the rest of this paper, we will always focus on finding a finest SBD.
We write
\begin{equation}
    T=\mathcal{SBD}(A^{(1)},A^{(2)},...,A^{(M)})
\end{equation}
to indicate that the following transformation yields
\begin{equation}
    T^{-1} A^{(k)} T=B^{(k)}, \quad k=1,...,M,
\end{equation}
where all the matrices $B^{(k)}$, $k=1,...,M$ share the same finest block diagonal form,
\begin{equation}
    B^{(k)}=\bigoplus_j B_j^{(k)},
\end{equation}
with the blocks $B_j^{(k)}$ all having the same sizes $\beta_k$ for $k=1,...,M$ and not being further reducible {by a simultaneous transformation.}

The method for
calculating a finest SBD transformation proposed in \cite{Maehara2011} requires two main steps: (i) Finding a matrix $P$ that commutes with a set of matrices, and (ii) calculating the transformation matrix $T$ formed of the eigenvectors of the matrix $P$.

\subsection{Matrix P}\label{sec:P}
In order to study stability of the cluster synchronization solution, we need to simultaneously block diagonalize the set of $C+1$ matrices $\lbrace A, E_1, E_2, \cdots, E_C \rbrace$ where the matrix $A$ is the adjacency matrix of the network and each $E_i$ is the binary and diagonal cluster indicator matrix \cite{Zhang2020}. Once we find $T=\mathcal{SBD}(A, E_1, E_2,..., E_C)$, the matrices $A$ and $E_i$ are transformed as follows,
\begin{subequations}\label{eq:directsumLR}
\begin{equation}
    T^{-1} A T=  B=\oplus_{k=1}^{r} \hat{B}^k, \label{eq:directsumL}
\end{equation}
\begin{equation}
    T^{-1} E_i T=  J_{i}=\oplus_{k=1}^{r} \hat{J}^k_i \quad i = 1, \cdots, C, \label{eq:directsumR}
\end{equation}
\end{subequations}
where $B$ is the transformed matrix $A$, $J_i$ is the transformed matrix $E_i$, and the blocks of the $C+1$-tuple $(\hat{B}^k,\hat{J}^k_1,....,\hat{J}^k_C)$ have the same dimensions $\beta_k$, $k=1,...,r$, $\sum_{k=1}^r \beta_k=N$. In what follows, we will refer to $(\hat{B}^k,\hat{J}^k_1,....,\hat{J}^k_C)$ as a block tuple.

Without loss of generality, we assume that the network nodes are ordered so that the first $n_1$ nodes are the ones in cluster $\mathcal{C}_1$, followed by the $n_2$ nodes in cluster $\mathcal{C}_2$, and so on, and the last $n_C$ nodes are the ones in cluster $\mathcal{C}_C$.
Then the matrices $E_k$ are in the following form,
\begin{equation}\label{E_i}
    E_1=\begin{pmatrix}
  I_{n_1} & 0\\ 
  0 & {0}_{N-n_{1}}
    \end{pmatrix}
    \quad
        E_2=\begin{pmatrix}
  {0}_{n_{1}} & 0 & 0\\ 
  0 & I_{n_2} & 0\\
  0 & 0 & {0}_{N-(n_{1}+n_{2})}
    \end{pmatrix}
    \quad
    \cdots
    \quad
    E_C=\begin{pmatrix}
  {0}_{N-n_{C}} & 0\\ 
  0 & I_{n_C}
    \end{pmatrix}
\end{equation}
where $I_{n_i}$ is the identity matrix of size $n_{i}$, and $0_{n_i}$ is the zero matrix of size $n_{i}$.

\begin{lemma}\label{Lem_P}
Any matrix $P$ that commutes with the set of matrices $E_i$ in Eq. \eqref{E_i} has the following block-diagonal structure,
\begin{equation}\label{Pblock}
P=\begin{pmatrix} 
  P_{1} & 0 & \cdots & 0\\
  0 & P_{2} & \cdots & 0\\
  \vdots & \vdots & \ddots & \vdots\\
  0 & 0 & \cdots & P_{C}
\end{pmatrix}
\end{equation}
where the block $P_{1}$ has dimension $n_{1}$, the block $P_2$ has dimension $n_{2}$, and so on.
\end{lemma}
\begin{proof}
The product $PE_{i}$ must have columns of zeros in the same place as $E_{i}$, so if $P$ is to commute with $E_{i}$ then $E_{i}P$ must also have these columns as zero.  Because $E_{i}$ has an identity block in its non-zero columns, this forces $P$ to have zeros in the intersection rows where $E_{i}$ has its identity block and the columns where $E_{i}$ is zero. Placing all such zeros as necessary for each of the $E_{i}$ gives the structure of $P$ as claimed.
\end{proof}

Following Lemma \ref{Lem_P}, we can see that the transformation matrix $T$, which has the eigenvectors of the matrix $P$ for its columns,  must also have the same block-diagonal structure as the matrix $P$,
\begin{equation}\label{Tblock}
T=\begin{pmatrix} 
  T_{1} & 0 & \cdots & 0\\
  0 & T_{2} & \cdots & 0\\
  \vdots & \vdots & \ddots & \vdots\\
  0 & 0 & \cdots & T_{C}
\end{pmatrix}
\end{equation}
where $T_{1}$ is the matrix of eigenvectors of $P_{1}$, $T_{2}$ is the matrix of eigenvectors of $P_{2}$, and so on.

\begin{remark}
A trivial consequence of Eq.\ \eqref{Tblock} is that as the matrix $T$ is orthogonal, each one of the block matrices $T_k$ is also orthogonal, $T_k^T=T_k^{-1}$.
\end{remark}

\begin{remark} \label{interpretation}
The particular block-diagonal structure of the matrix $T$, where each block $T_k$ corresponds to cluster $\mathcal{C}_k$ has important consequences. Each {independent} block obtained from application of the SBD transformation will only produce linear combinations of the states of nodes from the same cluster (same color). This implies that to each block corresponds a colored subnetwork (corresponding to the block), see Definition 2. It also follows that each block describes stability of either one cluster or a set of intertwined clusters \cite{Pecora2014,siddique2018symmetry}.
\end{remark}

\begin{remark} \label{interpretation2}
An additional benefit of our matrix $T$ is that each one of its rows is associated to one and only one of the clusters, in the sense that all the entries of that row that do not correspond to the nodes of that cluster are zero. This is important as it indicates the particular way in which the cluster is broken if the Lyapunov exponent of the corresponding block of the matrix $B$ is positive \cite{siddique2018symmetry}, i.e., entries that are the same (different) correspond to nodes that remain (do not remain) synchronized after the breaking. 
\end{remark}

\begin{lemma}
Application of the block diagonal transformation matrix $T$ to each cluster indicator matrix $E_i$ ensures that $J_i = E_i$ for $i = 1, 2, \cdots, C$. In other words, each $E_i$ gets mapped back to itself using the transformation matrix $T$ of Eq. \eqref{Tblock}.
\end{lemma}
\begin{proof}
By considering the matrices $\lbrace E_1, E_2, \cdots, E_C \rbrace$ of Eq.\ \eqref{E_i} and the block diagonal transformation matrix $T$ of Eq.\ \eqref{Tblock} and by using the fact that a block-diagonal matrix can be inverted block by block, we have $T^{-1} E_i T = E_i$ for $i = 1, 2, \cdots, C$. For instance, for $E_1$ we have:
\begin{equation}
T^{-1} E_1 T=
\begin{pmatrix} 
  T^{-1}_{1} & 0 & \cdots & 0\\
  0 & T^{-1}_{2} & \cdots & 0\\
  \vdots & \vdots & \ddots & \vdots\\
  0 & 0 & \cdots & T^{-1}_{C}
\end{pmatrix}
\begin{pmatrix} 
  I_{n1} & 0 & \cdots & 0\\
  0 & 0 & \cdots & 0\\
  \vdots & \vdots & \ddots & \vdots\\
  0 & 0 & \cdots & 0
\end{pmatrix}
\begin{pmatrix} 
  T_{1} & 0 & \cdots & 0\\
  0 & T_{2} & \cdots & 0\\
  \vdots & \vdots & \ddots & \vdots\\
  0 & 0 & \cdots & T_{C}
\end{pmatrix}=
\begin{pmatrix} 
  I_{c1} & 0 & \cdots & 0\\
  0 & 0 & \cdots & 0\\
  \vdots & \vdots & \ddots & \vdots\\
  0 & 0 & \cdots & 0
\end{pmatrix}
\label{proof BT}
\end{equation}
\end{proof}

\begin{lemma} \label{lemmasame}
Each block $T_i$ of the matrix $T$ has a column with entries that are all the same.
\end{lemma}
\begin{proof}
According to Lemma \ref{Lem_P} and Eq.\ \eqref{Tblock}, we know that both the matrices $P$ and $T$ are block-diagonal in $C$ blocks, where each block corresponds to one cluster. Also, according to Ref.\ \cite{sanchez2020exploiting} the adjacency matrix $A$ corresponding to a network with $C$ equitable clusters, has $C$ `non-redundant'
eigenvectors $\mathbf{v}_k$, each one associated with a cluster $\mathcal{C}_k$, $k=1,...,C$ whose entries ${\mathbf{v}_k}_i$ are all the same for $i \in \mathcal{C}_k$ and zero otherwise. From the facts that the matrix $P$ commutes with $A$ and has  block-diagonal structure (Eq. $\eqref{Pblock}$) it follows that the matrix $P$ will also have eigenvectors $\mathbf{v}_k$, $k=1,...,C$.  Thus each block $T_k$ associated with cluster $\mathcal{C}_k$ has a column with entries that are all the same and equal to ${n_k}^{-1/2}$.
\end{proof}


\begin{remark}
The transformation matrix $T$ has two properties in common with the transformation introduced in Refs.\ \cite{Pecora2014,della2020symmetries}: one is the block diagonal structure \eqref{Tblock} and the other one is the property (Lemma \ref{lemmasame}) that each block has a column with entries that are all the same.  Thus the two transformations are similar, with the following differences: (1) the transformation matrix $T$ is not based on calculation of the network symmetries and can be successfully applied to the case of equitable clusters and (2) the transformation matrix $T$ is faster to compute.
\end{remark}

\begin{lemma} \label{Q}
One of the block tuples resulting from the SBD transformation of the set $\{A,E_1,E_2,...,E_C\}$ is $C$-dimensional and corresponds to the quotient network dynamics.
\end{lemma}
\begin{proof}
As the $T$ matrix is orthogonal $T^{-1}=T^T$. From the matrix $T^T$ it is possible to extract $C$ rows, where each row $i=1,...,C$ includes the column of the block $T_i$ with entries that are all the same and equal to $n_i^{-1/2}$. We call $\Delta$ the matrix obtained by stacking together these $C$ rows. We note that the so-constructed matrix $\Delta=(O^T O)^{-1/2} O^T$. Then the $C \times C$-dimensional block,
\begin{equation}
    \hat{B}^1=\Delta A \Delta^T=(O^T O)^{1/2} Q(O^T O)^{-1/2},
\end{equation}
from which we see that the two matrices $\hat{B}^1$ and $Q$ are similar.
A similar proof can be found in \cite{klickstein2019symmetry}. 

\end{proof}

\begin{remark}
Lemma \ref{Q} is useful as it allows us to identify one block-tuple that is associated with dynamics parallel to {the} cluster synchronization manifold (the dynamics of the quotient network.)
We use the label $k=1$ to indicate this block-tuple $(\hat{B}^1,\hat{J}^1_1,....,\hat{J}^1_C)$. We use the labels $k=2,...,C$ to indicate the block-tuples associated with dynamics transverse to the synchronization manifold. The transverse block-tuples determine stability of the cluster synchronous solution \cite{Pecora2014}.
\end{remark}

\begin{lemma}\label{lemmaPi}
The sub-matrices $P_i \quad i= 1, 2, \cdots, C$ can be found by computing the null subspace of the $\sum_i n_{i}^2 \times \sum_i n_{i}^2$-dimensional matrix $S^T S$ defined below.
\end{lemma}

\begin{proof}
By rewriting the adjacency matrix $A$ as:
\begin{equation}
    \begin{pmatrix}
      A_{11} & A_{12} & \cdots & A_{1C}\\
      A_{21} & A_{22} & \cdots & A_{2C}\\
      \vdots & \vdots & \ddots & \vdots\\
      A_{C1} & A_{C2} & \cdots & A_{CC}\\
    \end{pmatrix}
\end{equation}
the commutation equation $P{A}={A}P$ becomes
\begin{subequations}\label{AP}
\begin{equation}\label{APa}
\begin{pmatrix} 
  P_{1} & 0 & \cdots & 0\\
  0 & P_{2} & \cdots & 0\\
  \vdots & \vdots & \ddots & \vdots\\
  0 & 0 & \cdots & P_{C}
\end{pmatrix}    \begin{pmatrix}
      A_{11} & A_{12} & \cdots & A_{1C}\\
      A_{21} & A_{22} & \cdots & A_{2C}\\
      \vdots & \vdots & \ddots & \vdots\\
      A_{C1} & A_{C2} & \cdots & A_{CC}\\
    \end{pmatrix}=    \begin{pmatrix}
      A_{11} & A_{12} & \cdots & A_{1C}\\
      A_{21} & A_{22} & \cdots & A_{2C}\\
      \vdots & \vdots & \ddots & \vdots\\
      A_{C1} & A_{C2} & \cdots & A_{CC}\\
    \end{pmatrix} \begin{pmatrix} 
  P_{1} & 0 & \cdots & 0\\
  0 & P_{2} & \cdots & 0\\
  \vdots & \vdots & \ddots & \vdots\\
  0 & 0 & \cdots & P_{C}
\end{pmatrix}
\end{equation}
\begin{equation}\label{APb}
   \begin{pmatrix}
      P_1 A_{11} & P_1 A_{12} & \cdots & P_1 A_{1C}\\
      P_2 A_{21} & P_2 A_{22} & \cdots & P_2 A_{2C}\\
      \vdots & \vdots & \ddots & \vdots\\
      P_C A_{C1} & P_C A_{C2} & \cdots & P_C A_{CC}\\
    \end{pmatrix}=\begin{pmatrix}
      A_{11}P_1 & A_{12}P_2 & \cdots & A_{1C}P_C\\
      A_{21}P_1 & A_{22}P_2 & \cdots & A_{2C}P_C\\
      \vdots & \vdots & \ddots & \vdots\\
      A_{C1}P_1 & A_{C2}P_2 & \cdots & A_{CC}P_C\\
    \end{pmatrix}.
\end{equation}
\label{Eq13}
\end{subequations}

Equation \eqref{APb} corresponds to the following $C^2$ equations which should all be simultaneously solved:
\begin{equation}\label{nPA}
    P_i A_{ij} - A_{ij}P_j=0_{n_{i}, n_{j}} \quad i,j=1, 2, \cdots, C
\end{equation}
where $n_{i}$ is the dimension of the sub-matrix $P_i$.

Define the function $\text{vec}:\mathbb{R}^{n \times m} \mapsto \mathbb{R}^{nm}$ that maps a matrix to a vector by stacking the columns of the matrix.
We obtain the set of $C^2$ equations,
 \begin{equation}\label{setAP}
     (A_{ij}^{T} \otimes I_{n_{pi}})vec(P_i)-(I_{n_{pi}} \otimes A_{ij})vec(P_j)=0_{n_{i}, n_{j}}
 \end{equation}
Accordingly, Eq.\ \eqref{setAP} can be expressed as two linear systems of equations in the matrices,
\begin{subequations}\label{eq:nullspaces}
\begin{equation}\label{eq:nullspacesa}
    S_1 = \begin{pmatrix}
       \bar{A}_1 & 0 & \cdots & 0\\
       0 & \bar{A}_2 & \cdots & 0\\
       \vdots & \vdots & \ddots & \vdots\\
       0 & 0 & \cdots & \bar{A}_C
    \end{pmatrix} 
\end{equation}
\begin{equation}
    S_2=\begin{pmatrix}\label{eq:nullspacesb}
       A_{12}^{T} \otimes I_{n_{1}} & -I_{n_{2}}\otimes A_{12} & 0 & 0 & \cdots & 0 & 0\\
       -I_{n_{1}} \otimes A_{21} & A_{21}^{T}\otimes I_{n_{2}} & 0 & 0 & \cdots & 0 & 0\\
       A_{13}^{T} \otimes I_{n_{1}} & 0 & -I_{n_{3}}\otimes A_{13} & 0 &\cdots & 0 & 0\\
       -I_{n_{1}} \otimes A_{31} & 0 & A_{31}^{T}\otimes I_{n_{3}} & 0 &\cdots & 0 & 0\\
       \vdots & \vdots & \vdots & \vdots & \ddots & \vdots & \vdots\\
       A_{1C}^{T} \otimes I_{n_{1}} & 0 & 0 & 0 & \cdots & 0 &-I_{n_{C}}\otimes A_{1C}\\
        -I_{n_{1}}\otimes A_{C1} & 0 & 0 & 0 & \cdots & 0 &A_{C1}^{T}\otimes I_{n_{C}}\\
       0 & A_{23}^T \otimes I_{n_{2}} & -I_{n_{3}} \otimes A_{23} & 0 & \cdots & 0 & 0\\
       0 & -I_{n_{2}} \otimes A_{32} & A_{32}^T \otimes I_{n_{3}} & 0 & \cdots & 0 & 0\\
       \vdots & \vdots & \vdots & \vdots & \ddots & \vdots & \vdots\\
       0 & A_{2C}^T \otimes I_{n_{2}} & 0 & 0 & \cdots & 0 & -I_{n_{C}} \otimes A_{2C}\\
       0 & -I_{n_{2}} \otimes A_{C2} & 0 & 0 & \cdots & 0 & A_{C2}^T \otimes I_{n_{C}}\\
       \vdots & \vdots & \vdots & \vdots & \ddots & \vdots & \vdots\\
       0 & 0 & 0 & 0 & \cdots & A_{C-1,C}^T \otimes I_{n_{(C-1)}} & -I_{n_{C}} \otimes A_{C-1,C}\\
       0 & 0 & 0 & 0 & \cdots & -I_{n_{(C-1)}} \otimes A_{C,C-1} & A_{C,C-1}^T \otimes I_{n_{C}}
    \end{pmatrix}
\end{equation}
\end{subequations}
where $\bar{A}_{i}=(A_{ii}^{T} \otimes I_{n_{i}})-(I_{n_{i}} \otimes A_{ii})$. 
By stacking Eq. \eqref{eq:nullspacesa} and \eqref{eq:nullspacesb} together as
$S=\begin{bmatrix} S_1^T & S_2^T \end{bmatrix}^T$, we search for a vector $\boldsymbol{p} = \begin{bmatrix}
   \text{vec}(P_1)^T & \text{vec}(P_2)^T & \cdots & \text{vec}(P_C)^T
\end{bmatrix}^T$ that lies in $\mathcal{N}(S)$, the nullspace of $S$.
The matrix $S \in \mathbb{R}^{N_r \times N_c}$ has number of rows $N_r=\sum_{i=1}^{C}\sum_{j=1}^{C}n_{i} n_{j}$ and number of columns $N_c=\sum_{i=1}^{C} n_{i}^2$.
By using the property of nullspaces that $\mathcal{N}(S) = \mathcal{N}(S^TS)$, we instead look for a vector $\boldsymbol{p}$ that lies in $\mathcal{N}(S^TS)$,
\begin{equation}
    S^T S \boldsymbol{p} = \boldsymbol{0}_{N_c},
\end{equation}
where $S^TS \in \mathbb{R}^{N_c \times N_c}$.
\end{proof}

\begin{remark}
It is worth mentioning that the original method \cite{Maehara2011} requires finding a vector in the nullspace of a $N^2 \times N^2$-dimensional matrix while the method derived here requires finding a vector in the nullspace of a $\sum_i n_{i}^2 \times \sum_i n_{i}^2$-dimensional matrix where $\sum_i n_{i}^2 $ is not greater than $N^2$.
\begin{equation}
    n^2=(\sum_{i}n_{i})^2 \geq \sum_{i} n_{i}^2
\end{equation}
Equality only occurs if $C = 1$ while the minimum of $\sum_i n_{i}^2$ is achieved when $C = N$ (so $n_{i} = 1$).
\end{remark}

By applying $T$ to Eq.\ \eqref{z}, we obtain,
\begin{equation}
\begin{array}{lcl}
\dot{\pmb{\eta}}(t) = \left[(T^{-1}\sum_{k=1}^{C}E_{k} T) \otimes D\bF(\bs_{k}(t)) + (T^{-1}\sum_{k=1}^{C}A E_{k} T) \otimes D\bH (\bs_{k}(t))\right] \pmb{\eta}(t),
\end{array}
\label{Eq10}
\end{equation}
 where $\pmb{\eta}(t) = (T^{-1} \otimes I_{m}) \delta \bx(t)$. Eq.\ \eqref{Eq10} can be rewritten,
 \begin{equation}
     \dot{\pmb{\eta}}(t)= \left[\sum_{k=1}^C J_{k} \otimes D\bF(\bs_{k}(t)) + \sum_{k=1}^C B J_{k} \otimes D\bH(\bs_{k}(t))\right]{\pmb \eta}(t),
     \label{Eq11}
 \end{equation}
 which can be decoupled into $r$ independent equations of smaller dimensions, in the blocks of the block diagonal matrices $J_k$ and $B J_k$.
 

 \subsection{Importance of the canonical transformation}
 We call an SBD transformation of the set of $C+1$ matrices $\{A,E_1,E_2,...,E_C\}$ \textit{`canonical'} if the transformation maps the matrices $E_1,E_2,...,E_C$ back to themselves. There are at least two main strengths of a canonical SBD transformation: (i) the parametrization associated with the SBD transformation, 
 and (ii) the interpretation of the matrices resulting from this transformation. We briefly comment on both {strengths} below. Consider that the blocks obtained by the SBD transformation of the set of $C+1$ matrices $\{A,E_1,E_2,...,E_C\}$ have sizes $\beta_k$, $k=1,...,r$, $\sum_{k=1}^r \beta_k=N$.  Then the number of nonzero entries that parametrize the matrices after the transformation is  equal to $p_1=(C+1) \sum_{k=1}^r \beta_k(\beta_k+1)/2$. In the case of a canonical SBD transformation the number of nonzero entries that parametrize the matrices after the transformation is equal to $p_2=\sum_{k=1}^r \beta_k(\beta_k+1)/2$, 
 $p_2<p_1$, taking into account that the matrices $J_1,J_2,...,J_C$ are known `a priori'. Also in terms of interpretation, the diagonal matrices $J_1,J_2,...,J_C$ with either zeros or ones on the main diagonal indicate that each network node after the transformation belongs to one and only one cluster, i.e., node $i$ belongs (does not belong) to cluster $k$ if the entry $(i,i)$ of matrix $J_k$ is one (zero). Therefore, one can still say that the resulting `nodes' have a color (corresponding to the cluster they belong to) while node-node interactions are limited to the transformed matrix $B$, see also remarks \ref{interpretation} and \ref{interpretation2}. 
 This is different from the case of a non-canonical transformation where the resulting `nodes' do not necessarily have a color (i.e., they do not correspond to {a single cluster} of the original network) and there are in general $C+1$  `layers of connectivity' between these nodes.
 
\begin{figure}[H]
\centering
  \includegraphics[scale=.55]{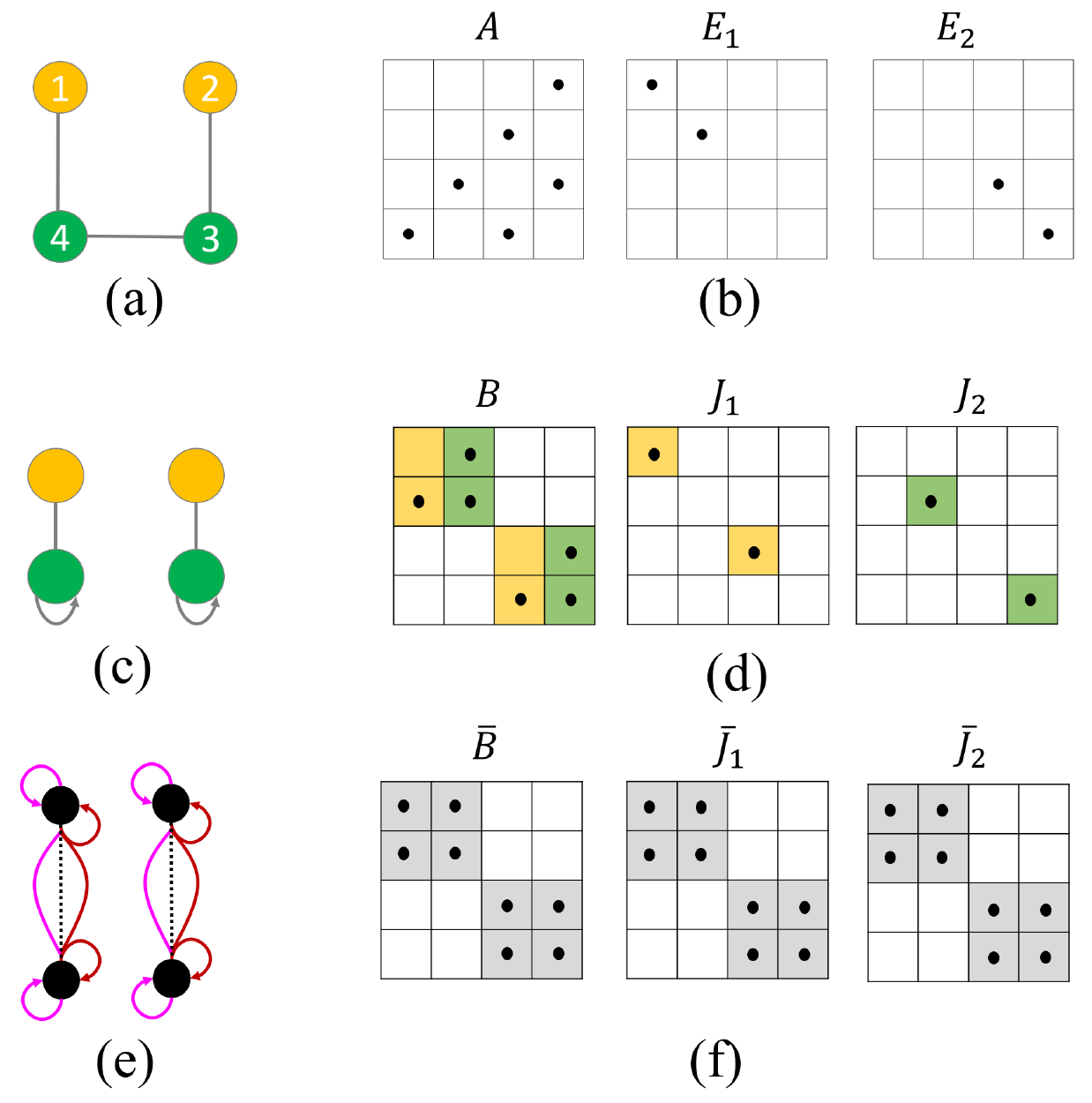}
\caption{(a) An $N=4$ node network with $C=2$ clusters. Nodes are color-coded according to the clusters to which they belong. (b) The adjacency matrix of the network ($A$) and the two cluster indicator matrices ($(E_1,E_2)$) are graphically shown, in which a nonzero entry is indicated as a black dot. (c) and (e) show the quotient and transverse sub-networks after application of the SBD transformation $T$ and $\tilde{T}$, respectively. Subnetwork nodes are colored based on the cluster with which they are associated. Black nodes indicate that they are not associated with only one cluster. In (e) connections with different colors represent coupling through the $\bar{B}$ subnetwork and the $\bar{J}_i$ subnetworks.  
(d) The set of matrices $\{ B, J_1, J_2 \}$ and (f) the set of matrices $\{ \bar{B}, \bar{J}_{1}, \bar{J}_{2}\}$ 
obtained by using the transformation $T$ and $\tilde{T}$, respectively. The background  color of each matrix entry indicates the cluster about which the dynamics is linearized. A gray background color indicates that the dynamics is linearized about multiple clusters.  }
\label{4n}
\end{figure}

 The SBD transformation described in Sec.\ \ref{sec:P} is canonical while the SBD transformation proposed in \cite{Zhang2020} is not. For example, consider the $N=4$-dimensional network with $C=2$ clusters, shown in Fig.\ \ref{4n} (a).  The nodes are colored according to the cluster to which they belong.
 Figure \ref{4n} shows the results after applying a canonical (c, d) and a non-canonical \cite{Zhang2020} (e, f) SBD transformation. It can be seen that both transformations reduce the $4m$-dimensional stability problem of Eq.\ \eqref{z} into a $2m$-dimensional equation corresponding to the quotient dynamics and a $2m$-dimensional equation corresponding to the transverse dynamics. Only the latter is responsible for stability of the cluster synchronous solution.
 Equation \eqref{z} for the network of Fig.\ \ref{4n} is:
 \begin{equation}
 \begin{array}{cc}
\delta \dot{{\bf X}}=\begin{pmatrix}
        \delta\dot{\bx_1}\\
       \delta \dot{\bx_2}\\
    \delta \dot{\bx_3}\\
        \delta \dot{\bx_4}\\
        \end{pmatrix}=\begin{pmatrix}
           \textcolor{BurntOrange}{D\bF(\bs_1)} & 0 & 0 & 0\\
           0 & \textcolor{BurntOrange}{D\bF(\bs_1)} & 0 & 0\\
           0 & 0 & \textcolor{OliveGreen}{D\bF(\bs_2)} & 0\\
           0 & 0 & 0& \textcolor{OliveGreen}{D\bF(\bs_2)}\\
        \end{pmatrix}\delta {\bf X}\\
      +\begin{pmatrix}
           0 & 0 & 0& \textcolor{OliveGreen}{D\bH(\bs_2)}\\
           0 & 0 & \textcolor{OliveGreen}{D\bH(\bs_2)} & 0\\
           0 & \textcolor{BurntOrange}{DH(\bs_1)} & 0& \textcolor{OliveGreen}{D\bH(\bs_2)}\\
           \textcolor{BurntOrange}{D\bH(\bs_1)} & 0& \textcolor{OliveGreen}{D\bH(\bs_2)} & 0\\
        \end{pmatrix}\delta {\bf X}.
 \end{array}\label{xdelt}
 \end{equation}
 Using the canonical SBD transformation matrix $T$, Eq. \eqref{xdelt} becomes,
  \begin{equation}
 \begin{array}{cc}
 \dot{{\bf Y}}=T^{-1}\delta \dot{{\bf X}}=\frac{\sqrt{2}}{2} \begin{pmatrix}
        \delta \dot{\bx}_1 + \delta \dot{\bx}_2\\
        -\delta \dot{\bx}_3 - \delta \dot{\bx}_4\\
         \delta \dot{\bx}_1 - \delta \dot{\bx}_2\\
         \delta \dot{\bx}_3 - \delta \dot{\bx}_4\\
     \end{pmatrix}=
\begin{pmatrix}
           \textcolor{BurntOrange}{D\bF(\bs_1)} & 0 & 0 & 0\\
           0 & \textcolor{OliveGreen}{D\bF(\bs_2)} & 0 & 0\\
           0 & 0 & \textcolor{BurntOrange}{D\bF(\bs_1)} & 0\\
           0 & 0 & 0& \textcolor{OliveGreen}{D\bF(\bs_2)}\\
        \end{pmatrix}{\bf Y}\\
      +\begin{pmatrix}
           0 & - \textcolor{OliveGreen}{D\bH(\bs_2)} & 0& 0\\
           - \textcolor{BurntOrange}{D\bH(\bs_1)} & \textcolor{OliveGreen}{D\bH(\bs_2)} & 0 & 0\\
           0 & 0 & 0& \textcolor{OliveGreen}{D\bH(\bs_2)}\\
           0 & 0 & \textcolor{BurntOrange}{D\bH(\bs_1)} & - \textcolor{OliveGreen}{D\bH(\bs_2)}\\
        \end{pmatrix}{\bf Y},
 \end{array}\label{Ty}
 \end{equation}
 where the vectors ${\bf Y}=[{\bf Y}_1^T,{\bf Y}_2^T,{\bf Y}_3^T,{\bf Y}_4^T]^T$ in Eq. \eqref{Ty} is the transformed vector $\delta {\bf X}$ using canonical ($T$) transformation. 
 We see that the dynamics of each transformed variable ${\bf Y}_i$ is described by either one of the two Jacobians $D\bF(\bs_1)$ and $D\bF(\bs_2)$. This is due to the fact that the canonical transformation $T$ only produces linear combinations of nodes belonging to the same cluster (see Remark \ref{interpretation}.) For example, the yellow nodes in Fig.\ \ref{4n} (a)  get mapped back to other yellow nodes in Fig.\ \ref{4n} (c), and so on. These transformed nodes are then coupled through the connections of only one network corresponding to the resulting matrix $B$ (in two blocks.) 
 
 Using the non-canonical SBD transformation matrix $\tilde{T}$, Eq.\ \eqref{xdelt} becomes,
 \begin{equation}
{
 \begin{array}{ccc}
 \dot{{\bf Z}}=\tilde{T}^{-1}\delta \dot{{\bf X}}=
 \begin{pmatrix}
       -0.65 \delta \dot{\bx}_1 + 0.65 \delta \dot{\bx}_2 - 0.26 \delta \dot{\bx}_3 + 0.26 \delta \dot{\bx}_4\\
        - 0.26 \delta \dot{\bx}_1 + 0.26 \delta \dot{\bx}_2 + 0.65 \delta \dot{\bx}_3 - 0.65 \delta \dot{\bx}_4\\
        0.69 \delta \dot{\bx}_1 + 0.69 \delta \dot{\bx}_2 - 0.16 \delta \dot{\bx}_3 - 0.16 \delta \delta \dot{\bx}_4\\
        0.16 \delta \dot{\bx}_1 + 0.16 \delta \dot{\bx}_2 + 0.69 \delta \dot{\bx}_3 + 0.69 \delta \dot{\bx}_4\\
     \end{pmatrix}\\
     =
\begin{pmatrix}
           0.14 \textcolor{BurntOrange}{D\bF(\bs_1)}+ 0.86 \textcolor{OliveGreen}{D\bF(\bs_2)} & 0.35\textcolor{BurntOrange}{D\bF(\bs_1)} - 0.35 \textcolor{OliveGreen}{D\bF(\bs_2)} & 0 & 0\\
           0.35 \textcolor{BurntOrange}{D\bF(\bs_1)} - 0.35  \textcolor{OliveGreen}{D\bF(\bs_2)} & 0.86 \textcolor{BurntOrange}{D\bF(\bs_1)}+ 0.14 \textcolor{OliveGreen}{D\bF(\bs_2)} & 0 & 0\\
           0 & 0 & 0.05 \textcolor{BurntOrange}{D\bF(\bs_1)}+ 0.95 \textcolor{OliveGreen}{D\bF(\bs_2)} & 0.22 \textcolor{BurntOrange}{D\bF(\bs_1)}- 0.22\textcolor{OliveGreen}{D\bF(\bs_2)}\\
           0 & 0 & 0.22 \textcolor{BurntOrange}{D\bF(\bs_1)} -0.22 \textcolor{OliveGreen}{D\bF(\bs_2)} & 0.95 \textcolor{BurntOrange}{D\bF(\bs_1)}+ 0.05 \textcolor{OliveGreen}{D\bF(\bs_2)}\\
        \end{pmatrix}{\bf Z}\\
      +\begin{pmatrix}
           0.35 \textcolor{BurntOrange}{D\bH(\bs_1)} - 0.51 \textcolor{OliveGreen}{D\bH(\bs_2)} & 0.86 \textcolor{BurntOrange}{D\bH(\bs_1)}+0.20\textcolor{OliveGreen}{D\bH(\bs_2)} & 0 & 0\\
           -0.14 \textcolor{BurntOrange}{D\bH(\bs_1)}+ 1.2\textcolor{OliveGreen}{D\bH(\bs_2)} & -0.35 \textcolor{BurntOrange}{D\bH(\bs_1)} -0.49\textcolor{OliveGreen}{D\bH(\bs_2)} & 0 & 0\\
           0 & 0 & 0.22\textcolor{BurntOrange}{D\bH(\bs_1)}+1.17\textcolor{OliveGreen}{D\bH(\bs_2)} & 0.95\textcolor{BurntOrange}{D\bH(\bs_1)}-0.27\textcolor{OliveGreen}{D\bH(\bs_2)}\\
           0 & 0 & -0.05\textcolor{BurntOrange}{D\bH(\bs_1)}+0.72\textcolor{OliveGreen}{D\bH(\bs_2)} & -0.22\textcolor{BurntOrange}{D\bH(\bs_1)}-0.17\textcolor{OliveGreen}{D\bH(\bs_2)}\\
        \end{pmatrix}{\bf Z},
 \end{array}}
 \label{Tz_zm}
 \end{equation}
 where the vector ${\bf Z}=[{\bf Z}_1^T,{\bf Z}_2^T,{\bf Z}_3^T,{\bf Z}_4^T]^T$ in Eq.\ \eqref{Tz_zm} is the transformed vector $\delta {\bf X}$ using the non-canonical ($\tilde{T}$) SBD transformation. We see that the transformation matrix $\tilde{T}$ produces the same number of blocks with the same dimensions as the transformation matrix $T$.
 However, from Eq.\ \eqref{Tz_zm} and Fig.\ \ref{4n} we see that application of the non-canonical SBD transformation does not preserve the color of the nodes in the original network. Also, we see that the individual ${\bf Z}_i$ are coupled with one another through both the $D\bF$ and the $D\bH$ terms. This is shown in Fig.\ \ref{4n}(e) where the resulting nodes are colored black to indicate that they do not correspond to any individual color of the original network. Connections of different colors are used to indicate the different types of coupling.
 
\begin{figure}[H]
\centering
\includegraphics[scale=.55]{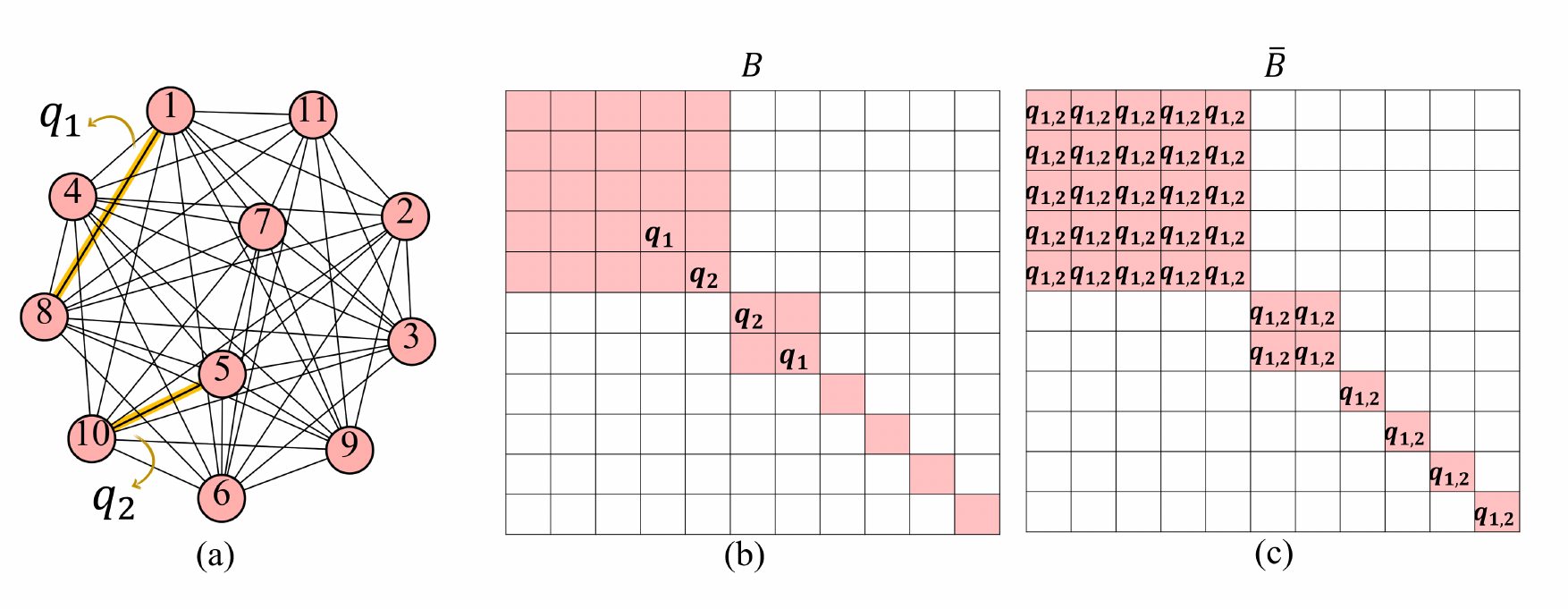}
  \caption{(a) $N=11$-dimensional network from Ref.\ \cite{Pecora2014} with two weighted edges.  The two edges between nodes $1$ and $8$ and nodes $5$ and $10$ are highlighted in yellow. We set $q_1=A_{1,8}=A_{8,1}$ and $q_2=A_{5,10}=A_{10,5}$ and parametrize the problem in terms of $q_{1}$ and $q_{2}$. (b) shows the matrix $B$ obtained from application of a canonical SBD transformation and (c) shows the matrix $\bar{B}$ obtained from application of a non-canonical SBD transformation. In both cases, we label with $q_1$ ($q_2$) the block entries that are affected by variations of the parameters $q_1$ ($q_2$.)}
\label{11n}
\end{figure}

 As {has been} stated before, the stability analysis of the cluster synchronization problem (Eq. \eqref{z}) is based on two main steps: the first one is the dimensionality reduction and the second one is the parametrization. Here, concerning the second step, we focus on the importance of a canonical SBD transformation in order to obtain a parametrization of the lower-dimensional problem in a minimal number of parameters. For example, consider the network from Ref.\ \cite{Pecora2014} shown in Fig.\ \ref{11n} (a). We  set $q_1=A_{1,8}=A_{8,1}$ and $q_2=A_{5,10}=A_{10,5}$ and parametrize the problem in terms of $q_1$ and $q_2$.  The two links between nodes $1$ and $8$ and nodes $5$ and $10$ are highlighted in yellow. Fig.\ \ref{11n} (b) and (c) show the block diagonal matrix $B$ and $\bar{B}$ obtained from application of the canonical ($T$) and non-canonical ($\tilde{T}$) transformation, respectively. In both cases, we indicate with $q_1$ ($q_2$) the block entries that are affected by variations of the parameters $q_{1}$ ($q_{2}$.) As we can see, changing $q_1$ and $q_2$ only affects a few entries of the block diagonal matrix in the case of the canonical transformation, but affects many more entries in the case of a non-canonical transformation. Looking at the transverse $2$-dimensional block in (b) we see that $q_1$ only affects one entry of this block and  $q_2$ only affects one other entry of that block. On the other hand, each entry of the $2$-dimensional block in (c) is affected by both $q_1$ and $q_2$.

 \section{Orbital and Equitable Partitions}
 
 We are now going to apply our approach to the case of a network for which the equitable and orbital clusters do not coincide. This network is from Ref.\ \cite{kudose2009}.  As for the other examples in this paper, the number and dimensions of the blocks will be the same as those obtained from the method in Ref.\ \cite{Zhang2020} but the structure of the blocks will be different. We will comment on the benefits of using a canonical transformation.

\begin{figure}[H]
\centering
  \includegraphics[scale=.55]{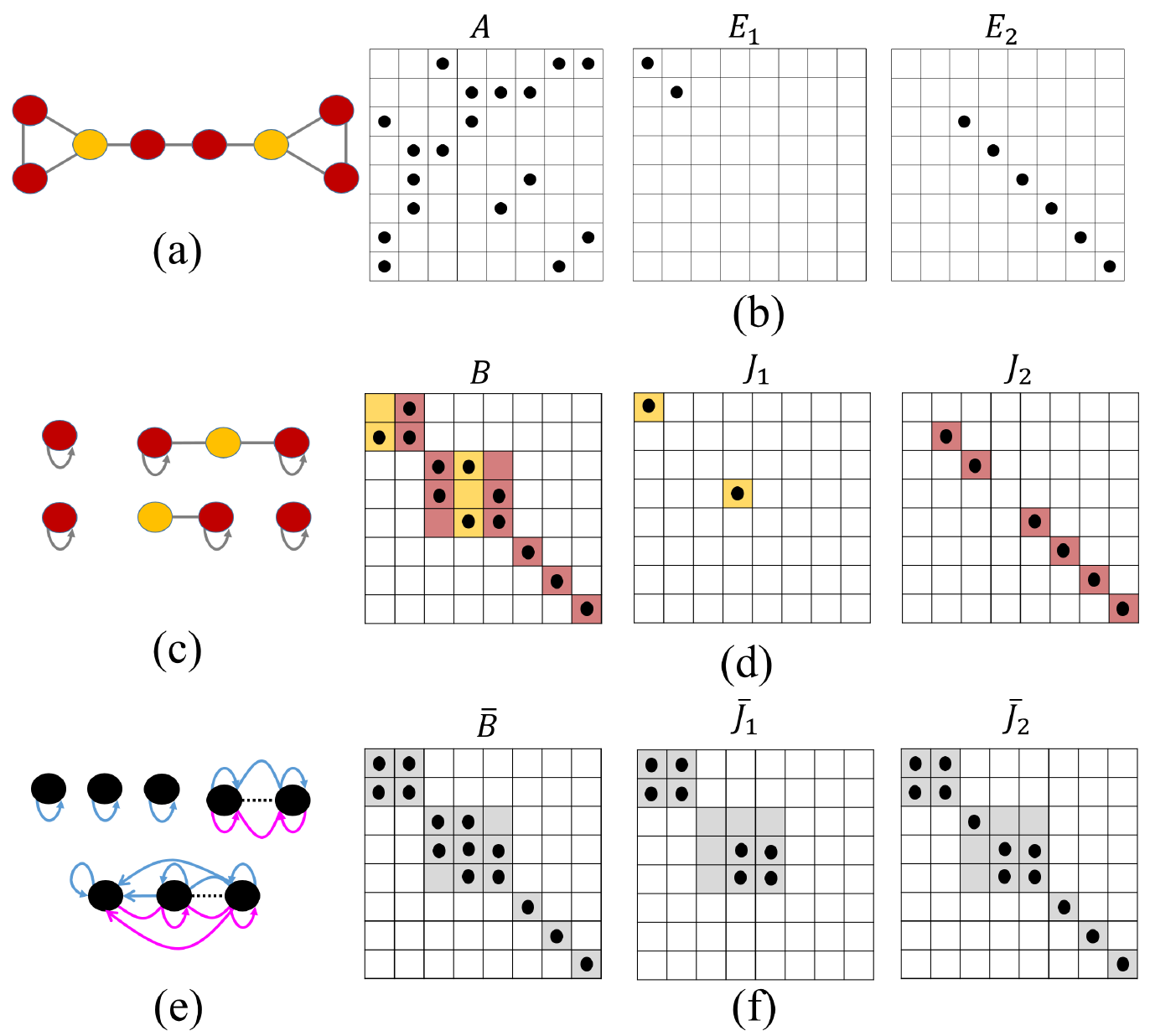}
  \caption{(a) An $N=8$-dimensional network with $C=2$ equitable clusters. Nodes are color-coded according to the clusters to which they belong. (b) The adjacency matrix of the network ($A$) and the two cluster indicator matrices ($(E_1,E_2)$) are graphically shown, in which a nonzero entry is indicated as a black dot. (c) and (e) show the quotient and transverse sub-networks after application of the SBD transformation $T$ and $\tilde{T}$, respectively. Subnetwork nodes are colored based on the cluster to which they are associated. Black nodes indicate that they are not associated with only one cluster. In (e) connections with different colors represent coupling through the $\bar{B}$ subnetwork and the $\bar{J}_i$ subnetworks.  
(d) The set of matrices $\{ B, J_1, J_2 \}$ and (f) the set of matrices $\{ \bar{B}, \bar{J}_{1}, \bar{J}_{2}\}$ obtained by using the transformation $T$ and $\tilde{T}$, respectively. The background color of each matrix entry indicates the cluster about which the dynamics is linearized. A gray background  color indicates that the dynamics is linearized about multiple clusters.}
\label{82}
\end{figure}

\begin{figure}[H]
\centering
\includegraphics[scale=.55]{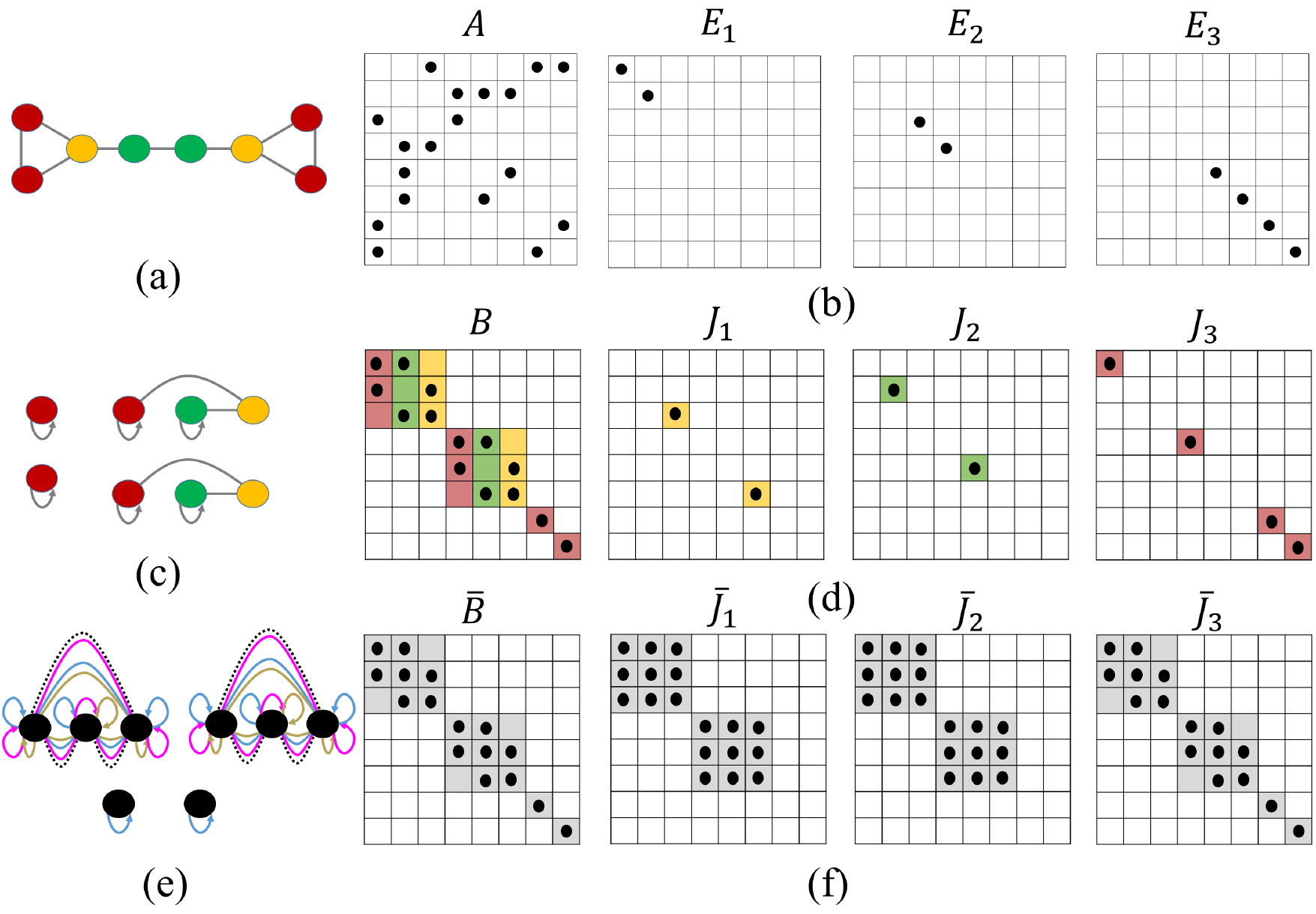}
  \caption{(a) An $N=8$-dimensional network with $C=3$ orbital clusters. Nodes are color-coded according to the clusters to which they belong. (b) The adjacency matrix of the network ($A$) and the three cluster indicator matrices ($(E_1,E_2,E_3)$) are graphically shown, in which a nonzero entry is indicated as a black dot. (c) and (e) show the quotient and transverse sub-networks after application of the SBD transformation $T$ and $\tilde{T}$, respectively. Subnetwork nodes are colored based on the cluster to which they are associated. Black nodes indicate that they are not associated with only one cluster. In (e) connections with different colors represent coupling through the $\bar{B}$ subnetwork and the $\bar{J}_i$ subnetworks.  
(d) The set of matrices $\{ B, J_1, J_2, J_3 \}$ and (f) the set of matrices $\{ \bar{B}, \bar{J}_{1}, \bar{J}_{2}, \bar{J}_3\}$ obtained by using the transformation $T$ and $\tilde{T}$, respectively. The background color of each matrix entry indicates the cluster about which the dynamics is linearized. A gray background color indicates that the dynamics is linearized about multiple clusters.}
\label{83}
\end{figure}

 
 We demonstrate the benefits of the canonical transformation with the $N=8$-dimensional network from Ref.\ \cite{kudose2009} shown in Fig. \ref{82} and \ref{83}. Figure \ref{82} is for the case of the network equitable partition with $C=2$ clusters and Fig. \ref{83} is for the case of the network orbital partition with $C=3$ clusters. 
 We compare the results after application of the canonical SBD transformation $T$ and the non-canonical SBD transformation $\tilde{T}$. Figure \ref{82} shows that in both cases the SBD leads to a total of $5$ blocks, one $2$-dimensional block corresponding to the quotient network, one $3$-dimensional block and three $1$-dimensional blocks, all corresponding to the transverse dynamics.
 
 An interesting case is that of the $3$-dimensional block, which is shown in panels (c) for the case of the canonical transformation and (e) for the case of the non-canonical transformation. In panel (c) the dynamics of each node is linearized about the dynamics of one and only one of the quotient network nodes (and so it retains the color of that node), while in panel (e) each node is associated to a linear combination of the dynamics of the quotient network nodes. Also the nodes in (e) are coupled through different types of connections, consistent with the non-diagonal structure of the $3$-dimensional blocks in (f).

Another example of a network for which the equitable and orbital clusters do not coincide from Ref.\ \cite{siddique2018symmetry} is presented in the Supplementary Information. For this other example, for both the case of the orbital and equitable partition, we obtain the same decomposition in blocks already presented in Ref.\ \cite{siddique2018symmetry}.

\newpage

\section{Real Networks Analysis}

{The ultimate goal of the study of stability of network cluster synchronization is to gain better understanding of real networks of interest. In particular, given a real network and a dynamics on its nodes, one would  like to know which patterns of cluster synchronization are possible for that network and how stable those patterns are.
It is thus important to consider application of the algorithm developed in this paper to real network topologies.}

In Table \ref{Table1} we apply both transformation $T$ and $\tilde{T}$ to block diagonalize the adjacency matrices of several real networks from 
the literature. For each network dataset, we include information on the number of nodes $N$, the number of edges $E$, the number of nontrivial clusters $N_{ntc}$ (clusters with more than one node \cite{Pecora2014}), the size of the largest equitable cluster $\max(|{n_{c}}|)$ and the {average runtime over $10$ numerical runs for calculation of the transformation matrices $T$ and $\tilde{T}$ (method from \cite{Zhang2020}).} All these networks are connected, undirected and unweighted (only the giant component was considered in the case of networks that are not connected.) For all the networks we have analyzed, the two transformation matrices $T$ and $\tilde{T}$ produced equivalent SBD decompositions, with the same number of blocks and of the same sizes. However, the computation time of our algorithm was 
(up to {six times}) faster. 
{It was previously shown that the code proposed in \cite{Zhang2020} is faster than the other codes proposed in Refs. \cite{Pecora2014,Maehara2011}(see Fig.\ 2 of Ref.\ \cite{Zhang2020}). Therefore, we only compare the run-time of our code  with that of  \cite{Zhang2020}}

\begin{center}
\begin{table}[h!]
\caption{Real networks analysis. $N$ is the number of nodes, $E$ the number of edges, $N_{ntc}$ is the number of nontrivial clusters, $\max(|{n_{c}}|)$ is the size of the largest equitable cluster. We include the average runtime in seconds for calculation of the transformation matrix $T$ and of the transformation matrix $\tilde{T}$ using the code from \cite{Zhang2020}.\\}
\begin{tabular}{ |c|c|c|c|c|c|c|c|} 
\hline
\shortstack{Name} & \shortstack{$\quad$ $N$ $\quad$} & \shortstack{$\quad$ $E$ $\quad$} & \shortstack{$N_{ntc}$} & \shortstack{$\max(|{n_{c}}|)$} & \shortstack{
{ Average} \\ Runtime \, for $\tilde{T}$\cite{Zhang2020}} & \shortstack{\\ 
{ Average}\\ Runtime\, \\ for ${T}$}\\
\hline
\shortstack{\\ ca-netscience: \\ Scientist Collaboration \\Network}\cite{nr-aaai15} & 379 & 914 & 70 & 6 & {25.0745} & {6.4077}\\
\hline
\shortstack{{Chilean Power Grid}\\Network\cite{kim2018depth,bhatta2021modal}} & 218 & 527 & 29 & 7 & {2.9198} & { 1.9107}\\
\hline
\shortstack{\, Power Grid Network\\of Western Germany}\cite{SciGRIDv0.2} & 491 & 665 & 43 & 5 & {41.7946} & {14.4780} \\
\hline
Metabolic Network\cite{nr-aaai15} & 453 & 2025 & 28 & 4 & {24.6675} & {12.6427} \\
 \hline
Us Airline\cite{pajek} & 332 & 2126 & 31 & 12 & {11.0195} & {5.8412} \\
 \hline
Erdos971\cite{nr-aaai15} & 429 & 1312 & 20 & 3 & {16.92} & {10.0446}\\
\hline
\shortstack{{celegans-dir:}\\
{Biological Network\cite{nr-aaai15}}} & {453} & {2025} & {28} & {4} & {23.5363} & {12.6916}\\
\hline
\shortstack{{bio-diseasome:}\\
{Biological Network\cite{nr-aaai15}}} & {516} & {1188} & {95} & {6} & {90.5763} & {13.9930}\\
\hline
\shortstack{{fb-forum:}\\
{social network\cite{nr-aaai15}}} & {899} & {7036} & {16} & {5} & {171.0608} & {65.6710}\\
\hline
\end{tabular}
\label{Table1}
\end{table}
\end{center}

\section{Conclusions}\label{s:conclusion}

In this paper we have studied cluster synchronization of networks and proposed a canonical transformation for  simultaneous block diagonalization of matrices
that we use to study stability of the cluster synchronous (CS) solution. Our approach presents several advantages as it allows us to: (1) decouple the stability problem into blocks of minimal dimensionality, while preserving physically meaningful information; (2) study stability of the CS solution for both the cases of orbital and equitable partitions of the network nodes and (3) obtain a parametrization of the problem in a minimal number of parameters.

When applied to several real network toplogies, our algorithm is faster than the one proposed in \cite{Zhang2020} and leads to a decomposition of the stability problem into a number of sub-problems that preserve key physical properties of the original system (such as the `color' of the nodes).  The main advantage of our canonical transformation is that it allows a parametrization of the stability problem in a small number of parameters.
In particular, we show how the stability analysis for 
different networks can be studied in terms of these parameters and how changing the coupling strengths of a subset of connections is reflected in `localized' variations of these parameters.  {With this paper, we provide a link to our code, which we hope will be used in conjunction with codes by other groups, such as e.g., \cite{Maehara2012Online} and \cite{Zhang2020}.}

\section{Supplementary Material}
In the supplementary material, we present other two examples of networks. The first example is a network with $N=6$ nodes. The second example with $N=10$ nodes is from \cite{siddique2018symmetry} and has different  orbital and equitable cluster partitions. For both the case of the orbital and of the equitable partition, we obtain the same decomposition in blocks previously found in Ref.\ \cite{siddique2018symmetry}.
\color{black}

\section*{Acknowledgement}
The authors thank Galen Novello for insightful conversations on the subject of $*$-algebra.

\section*{Author Declarations}
The authors have no conflicts to disclose.

\section*{Data Availability}
The data that supports the findings of this study are available within the article.

\section*{Code Availability}
The MATLAB code to compute the simultaneous block diagonalizations for the cluster synchronization examples shown in this paper can be accessed from its \textcolor{blue}{ \href{https://github.com/SPanahi/Clustered_SBD}{Github repo}}.

 \newcommand{\noop}[1]{}

\end{document}